\documentclass[12pt,preprint]{aastex}

\newcommand{\bdv}[1]{\mbox{\boldmath$#1$}}

\def\au{{\rm au}}

\def\kms{{\rm km}\,{\rm s}^{-1}}
\def\masyr{{\rm mas}\,{\rm yr}^{-1}}
\def\kpc{{\rm kpc}}
\def\mas{{\rm mas}}

\def\muas{\mu{\rm as}}

\def\rel{{\rm rel}}

\def\e{{\rm E}}
\def\bpi{{\bdv\pi}}
\def\bmu{{\bdv\mu}}

\def\bgamma{{\bdv\gamma}}

\def\bv{{\bf v}}

\begin{document}
\title{KMT-2018-BLG-1990Lb: A Nearby Jovian Planet From A Low-Cadence Microlensing Field}

\author{\textsc{
Yoon-Hyun Ryu$^{1}$, Kyu-Ha Hwang$^{1}$, Andrew Gould$^{1,4,5}$,
Michael D.Albrow$^{2}$, Sun-Ju Chung$^{1,3}$, Cheongho Han$^{6}$,
Youn Kil Jung$^{1}$, In-Gu Shin$^{7}$, Yossi Shvartzvald$^{8}$,
Jennifer C. Yee$^{7}$, Weicheng Zang$^{9}$, Sang-Mok Cha$^{1,10}$,
Dong-Jin Kim$^{1}$, Hyoun-Woo Kim$^{1}$, Seung-Lee Kim$^{1,3}$,
Chung-Uk Lee$^{1,3}$, Dong-Joo Lee$^{1}$, Yongseok Lee$^{1,10}$,
Byeong-Gon Park$^{1,3}$, Richard W. Pogge$^{5}$} }

\affil{$^{1}$Korea Astronomy and Space Science Institute, Daejon
34055, Republic of Korea}

\affil{$^{2}$University of Canterbury, Department of Physics and
Astronomy, Private Bag 4800, Christchurch 8020, New Zealand}

\affil{$^{3}$Korea University of Science and Technology, Daejeon
34113, Republic of Korea}

\affil{$^{4}$Max-Planck-Institute for Astronomy, K\"{o}nigstuhl 17,
69117 Heidelberg, Germany}

\affil{$^{5}$Department of Astronomy, Ohio State University, 140 W.
18th Ave., Columbus, OH 43210, USA}

\affil{$^{6}$Department of Physics, Chungbuk National University,
Cheongju 28644, Republic of Korea}

\affil{$^{7}$ Harvard-Smithsonian Center for Astrophysics, 60 Garden
St., Cambridge, MA 02138, USA}

\affil{$^{8}$IPAC, Mail Code 100-22, Caltech, 1200 E. California
Blvd., Pasadena, CA 91125, USA}


\affil{$^{9}$Physics Department and Tsinghua Centre for
Astrophysics, Tsinghua University, Beijing 100084, China}

\affil{$^{10}$School of Space Research, Kyung Hee University,
Yongin, Kyeonggi 17104, Republic of Korea}
\begin{abstract}
We report the discovery and characterization of KMT-2018-BLG-1990Lb,
a Jovian planet $(m_p=0.57_{-0.25}^{+0.79}\,M_J)$ orbiting a late M
dwarf $(M=0.14_{-0.06}^{+0.20}\,M_\odot)$, at a distance
$(D_L=1.23_{-0.43}^{+1.06}\,\kpc)$, and projected at $2.6\pm 0.6$
times the snow line distance, i.e., $a_{\rm snow}\equiv 2.7\,\au
(M/M_\odot)$, This is the second Jovian planet discovered by KMTNet
in its low cadence ($0.4\,{\rm hr}^{-1}$) fields, demonstrating that
this population will be well characterized based on survey-only
microlensing data.
\end{abstract}

\keywords{gravitational lensing: micro}

\section{{Introduction}
\label{sec:intro}}

The Korea Microlensing Telescope Network (KMTNet, \citealt{kmtnet})
monitors about 97 deg$^2$ toward the Galactic bulge from three
1.6m telescopes that are equipped with 4 deg$^2$ cameras,
in Chile (KMTC), South Africa (KMTS), and Australia (KMTA).  Its
primary goal is to find exoplanets via the anomalies that these
generate in microlensing events \citep{mao91}.

For many years, (beginning with the
second microlensing planet OGLE-2005-BLG-071Lb, \citealt{ob05071}), most
microlensing planets were discovered by intensive
follow-up observations of microlensing events that were alerted
by wide-field surveys. \citet{gouldloeb} had advocated such a
strategy because microlensing events, which have typical
Einstein timescales $t_\e\sim 20\,$days, can be discovered even
in very low-cadence surveys $\Gamma\la 1\,{\rm day}^{-1}$, whereas
the fleeting appearance of planetary anomalies requires much higher
cadence to detect and, more critically, to reasonably characterize
the planet.  In their scheme, the microlensing surveys could
cover very broad areas from even a single site, while narrow-angle
follow-up observations could be carried out around the clock at
much higher cadence on a handful of favorable targets.

By continuously monitoring a broad area at relatively high-cadence
from its three sites,
KMTNet aimed to simultaneously find microlensing events and
find and characterize the planetary anomalies within them,
without any followup observations (and hence without the necessity
of microlensing alerts).  In 2015, its first (commissioning) year of
observation, KMTNet narrowly focused on this strategy \citep{eventfinder}.
It observed only four fields, allowing a very high cadence
of $\Gamma = 6\,{\rm hr}^{-1}$, which as we will discuss immediately
below, was sufficient to detect planets down to about one Earth
mass\footnote{In fact, in 2015 KMTNet augmented this strategy with
very low-cadence (1--2 per day per observatory) observations
of 15 other fields in support of {\it Spitzer} microlensing.  However,
because these constituted $\la 10\%$ of all observations, they
did not significantly impact the basic strategy.}.

However, beginning in 2016, KMTNet developed a radically different
strategy, with a range of cadences $\Gamma=(4,1,0.4,0.2)\,{\rm hr}^{-1}$
covering
areas\footnote{This breakdown of cadences is
somewhat oversimplified.  See \citet{2016eventfinder} for a more detailed
summary.}
of (12,29,44,12) deg$^2$, respectively.  See Figure 12 of \citet{eventfinder}.
This change was motivated by a variety of goals, including support
for {\it Spitzer} microlensing
\citep{prop2013,prop2014,prop2015a,prop2015b,prop2016,prop2018},
which is an intrinsically wide-field experiment.  However, a major goal
was simply to find and characterize more planets over a much broader area.

The half-duration of a planetary perturbation is roughly the planetary
Einstein time
\begin{equation}
t_{\e,p} = {\theta_{\e,p}\over\mu_\rel} \simeq 1.25\,{\rm hr}
\biggl({m_p \over M_\oplus}\biggr)^{1/2}
\biggl({\pi_\rel \over 0.02\,\mas}\biggr)^{1/2}
\biggl({\mu_\rel \over 5\,\masyr}\biggr)^{-1},
\label{eqn:tep}
\end{equation}
where
\begin{equation}
\theta_{\e,p}\equiv \sqrt{q\over 1+q}\theta_\e;
\qquad
\theta_\e\equiv \sqrt{\kappa M\pi_\rel};
\qquad
\kappa\equiv {4 G\over c^2\au}\simeq 8.14\,{\mas\over M_\odot},
\label{eqn:thetae}
\end{equation}
$q\equiv m_p/M$ is the planet-host mass ratio, $\theta_\e$
is the Einstein radius, and $\pi_\rel$ and $\bmu_\rel$ are, respectively,
the lens-source relative parallax and proper motion.
Thus, assuming that about 10 data points are needed over the full
anomaly, the adopted cadences of $\Gamma=(4,1,0.4,0.2)\,{\rm hr}^{-1}$
should be sufficient to detect planets of mass
$m_p\sim(1,16,100,400)\,M_\oplus$, corresponding to ``Earth'', ``Neptune'',
``Saturn'' and ``Jupiter'' mass planets.  See also \citet{alertfinder}
and Figure 14 from \citet{henderson14}.
In effect, the revised strategy sacrificed sensitivity to Earth-mass
planets over 4 deg$^2$ in order to expand the total area by a factor 6,
including a 2.5 fold increase in the area sensitive to Neptunes.

Here we report on KMT-2018-BLG-1990Lb, a Jovian mass
planet lying in field BLG38, which is monitored by KMTNet at
$\Gamma=0.4\,{\rm hr}^{-1}$ and at a position that
was not monitored by other surveys.
It is the second KMT-only Jovian planet from a field with this
cadence, the first being KMT-2016-BLG-1397Lb from BLG31
\citep{kb161397}.  As such, it demonstrates the viability of a
wide-field survey for gas giant planets in accord with the KMTNet strategy.

\section{{Observations}
\label{sec:obs}}


KMT-2018-BLG-1990 is at (RA,Dec) = (17:53:44.49,$-22$:09:09.14),
corresponding to $(l,b)=(6.77,1.91)$, i.e., well out along the near side
of the Galactic bar.
As mentioned in Section~\ref{sec:intro},
it lies in KMTNet field BLG38, which has a nominal cadence of
$\Gamma=0.4\,{\rm hr}^{-1}$.  However, from the start of the season
through 25 June 2018, KMTNet followed a modified survey strategy in which
BLG38 continued to be observed at $\Gamma=0.4\,{\rm hr}^{-1}$ from KMTC,
but was observed at $\Gamma=0.3\,{\rm hr}^{-1}$ from KMTS and KMTA.
This period contained virtually the whole of the event.

The great majority of observations were carried out in the $I$ band,
but about 9\% were in $V$ band.  The primary purpose of the latter
was to determine the source color.
All reductions for this analysis were
conducted using variants of image subtraction
\citep{alard98}.  In particular, a variant of the \citet{wozniak2000}
difference image analysis (DIA) code was used for pipeline reductions
of the 500 million light curves that were searched for microlensing events.
Then pySIS \citep{albrow09} pipeline re-reductions were carried
out to further check whether the event is indeed microlensing, and finally
tender loving care (TLC) pySIS re-reductions were used to derive parameters.
A separate package, pyDIA \citep{pydia}, was used to construct
the color-magnitude diagram (CMD).

The event was discovered by applying the KMTNet event-finder algorithm
\citep{eventfinder} to the 2018 DIA light curves.
Very briefly, a machine review of these light curves
chose this event (among
100,130 candidates) for human review.  The DIA light curve
(which can be accessed at http://kmtnet.kasi.re.kr/ulens/)
is easily and securely identified as microlensing, but actually
displays no clear signature of an anomaly.  The anomaly was discovered
after running pipeline pySIS on the BLG38 candidates in order to
conduct a final review of these candidates.  The pipeline pySIS reduction
(also accessible at http://kmtnet.kasi.re.kr/ulens/) shows a
double-horned anomaly, characteristic of a planetary or binary caustic
crossing, which was immediately investigated and revealed to be
planetary.  By chance, this was the first 2018 event-finder
microlensing-event candidate that was examined manually.

\section{{Analysis}
\label{sec:analysis}}

\begin{figure}
\plotone{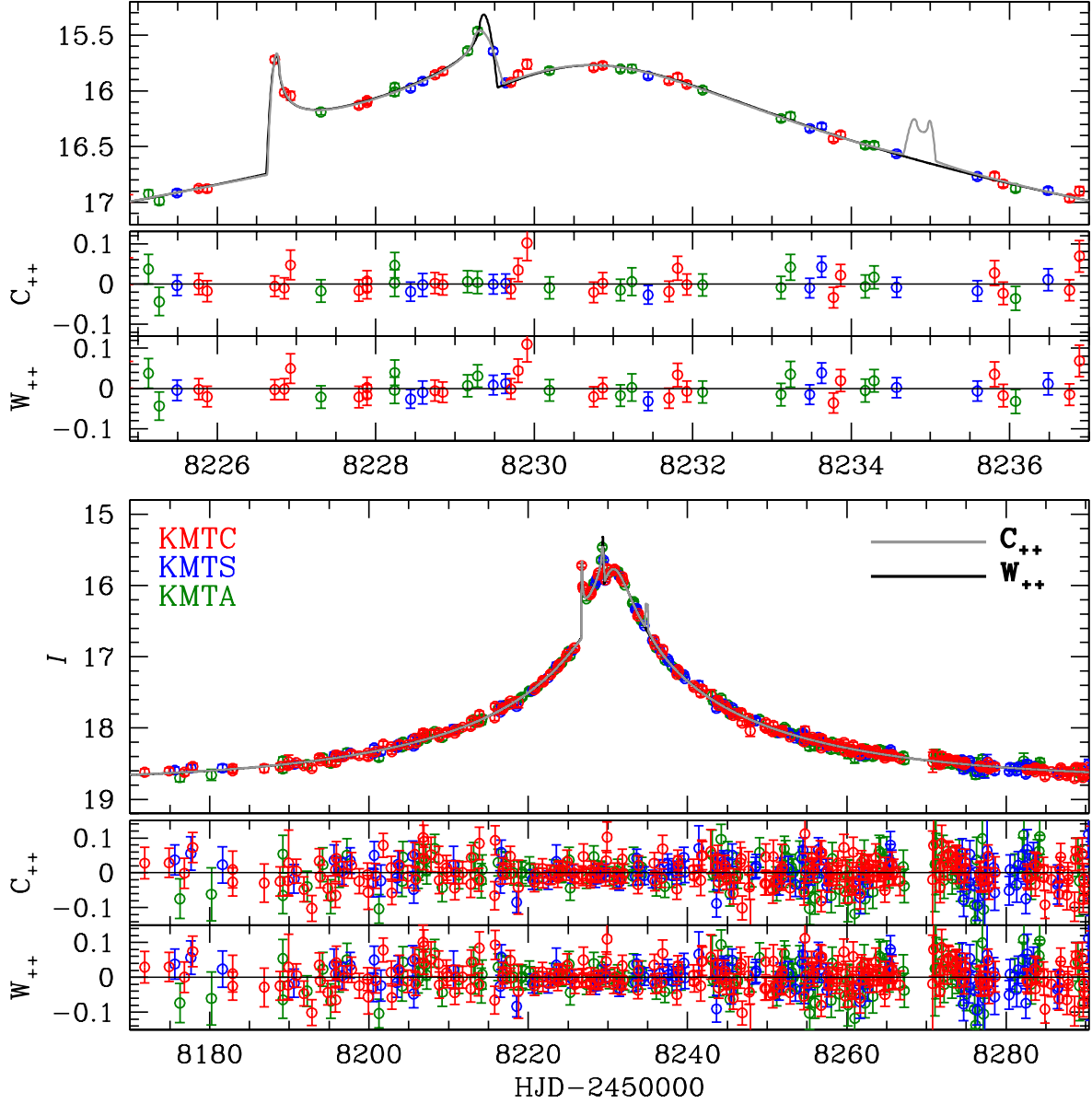}
\caption{Light curve of KMT-2018-BLG-1990 with data from the three
KMTNet observatories.  The curves show  the $C_{++}$ and $W_{++}$
models from Tables~\ref{tab:close} and \ref{tab:wide}, but the other
six degenerate models in these Tables
look extremely similar.
}
\label{fig:lc}
\end{figure}

Figure~\ref{fig:lc} shows the KMT-2018-BLG-1990 data together with
the best-fit model.  The light curve shows a roughly 65 hr anomaly
just before the peak of an otherwise normal \citet{pac86} point-lens
light curve.  According to the naive scaling given in Section~\ref{sec:intro},
this should correspond to an $m_p\sim 10^3\,M_\oplus$ companion.  It should
be noted that a broadly similar morphology can be generated when
a source transits the relatively small ``Chang-Refsdal'' caustic of
a wide ($s\gg 1$) or close ($s\ll 1$) binary with comparable-mass
components, followed by a cusp approach.  In this case, however,
one would expect that the peak would be offset from the center
of the light curve as defined by the wings.  Nevertheless, this
example serves as a caution that a thorough examination of parameter
space should be undertaken, even when the caustic geometry appears
``obvious''.

We proceed by a standard search for solutions defined by
seven non-linear parameters: $(t_0,u_0,t_\e,s,q,\alpha,\rho)$.
The first three are the \citet{pac86} parameters of the underlying
point-lens events: (time of closest approach, impact parameter
in units of $\theta_\e$,  Einstein timescale $t_\e=\theta_\e/\mu_\rel$).
The next three define the binary geometry: (separation in units
of $\theta_\e$, mass ratio, orientation of binary axis relative
to $\bmu_\rel$).  The last, $\rho=\theta_*/\theta_\e$, is the
source radius normalized to the Einstein radius.  In addition,
there are two flux parameters for each observatory $i$, i.e.,
the source flux $f_{s,i}$ and the blend flux $f_{b,i}$, so that the
model flux is given by $F_i(t) = f_{s,i}A(t) + f_{b,i}$, where
$A(t)$ is the model magnification.

We first conduct
a grid search, holding $(s,q)$ fixed on a grid of values and
allowing all other parameters to vary in a Monte Carlo Markov chain
(MCMC).  For the three \citet{pac86} parameters, we seed the
chains with the results from a point-lens fit (with the anomaly removed).
For $\alpha$, we seed with six values taken uniformly around the
unit circle.  Finally, we seed the normalized source radius
with $\rho=10^{-3}$.  The flux parameters
are determined by a linear fit to the data for each trial model.
We find that
there are two solutions, which are related by the so-called
close-wide degeneracy, which approximately takes $s\rightarrow s^{-1}$.
In fact, this degeneracy was originally derived in the limits
$s\gg 1$ or $s\ll 1$ \citep{griest98}, but sometimes holds when
$s\sim 1$ as well.

We then conduct a refined search on each of these two solutions by
allowing all seven geometric parameters to vary in the MCMC.  The
results are shown in Table~\ref{tab:ulens}.  Note that, in contrast to the
limiting cases from which this degeneracy is derived,  the
$s\rightarrow s^{-1}$ transformation does not preserve the value
of $q$.  Instead, $q(s<1)\sim 3.7\times 10^{-3}$, while
$q(s>1)\sim 6.2\times 10^{-3}$.

\subsection{Microlens Parallax: $\bpi_\e$}
\label{sec:bpie}

Next, we consider the microlens parallax effect.  A combination of
two facts strongly suggests that this effect will be measurable,
or at least strongly constrained.  First, the
timescale is relatively long, $t_\e\sim 45\,$days, implying that
Earth's velocity projected on the sky changes by $\Delta v\sim 35\,\kms$
during the $\sim 2 t_\e$ time interval when
the source is significantly magnified.

Second, as we will show in Section~\ref{sec:cmd}, $\theta_*\sim 1.2\,\muas$.
Combining this with the measured value of $\rho\sim 1.2\times 10^{-3}$
implies $\theta_\e=\theta_*/\rho\simeq 1.0\,\mas$.  The microlens
parallax vector is given by
\begin{equation}
\bpi_\e \equiv {\pi_\rel\over\theta_\e}\,{\bmu_\rel\over\mu_\rel},
\label{eqn:piedef}
\end{equation}
which immediately yields \citep{gould92,gould00},
\begin{equation}
\pi_\rel = \theta_\e\pi_\e;
\qquad
M = {\theta_\e\over\kappa\pi_\e}.
\label{eqn:massdist}
\end{equation}
Hence, $\theta_\e\simeq 1.0\,\mas$ implies $M\la 1\,M_\odot$ and
$\pi_\rel\la 0.13\,\mas$.  Otherwise the lens would be easily visible
in the blended light.  Then, from the limit on $\pi_\rel$,
we can also place limits on the amplitude of the lens-source projected velocity
$\tilde\bv\equiv (\au/t_\e)\bpi_\e/\pi_\e^2$,
\begin{equation}
\tilde v = {\au\over t_\e\pi_\e} = {\au\over\pi_\rel}\,{\theta_\e\over t_\e}
\la 290\,\kms.
\label{eqn:tildev}
\end{equation}
That is, during the event, the lens-source motion as observed from
Earth changes fractionally by at least $\Delta v/\tilde v\ga 12\%$.
We introduce two additional parameters $(\pi_{\e,N},\pi_{\e,E})$, which
are the components of $\bpi_\e$ in the Equatorial coordinate system.
Because the effects of Earth's orbital motion can be correlated
with the effects of lens orbital motion, it is essential to
simultaneously consider the latter, which we parameterize by
$(ds/dt,d\alpha/dt)$, the instantaneous changes in the separation and
orientation of the two components at $t_0$.
In fact, we find that the orbital parameters are only weakly constrained.  We
therefore restrict the MCMC trials by the condition $\beta<0.8$,
where $\beta$ is the ratio of projected kinetic to potential energy
\begin{equation}
\beta = \bigg|{\rm KE\over PE}\bigg|_\perp
= {\kappa M_\odot{\rm yr}^2\over 8\pi^2}{\pi_\e\over\theta_\e}
\gamma^2\biggl({s\over \pi_\e + \pi_s/\theta_\e}\biggr)^3;
\qquad \bgamma \equiv \biggl({ds/dt\over s},{d\alpha\over dt}\biggr),
\label{eqn:beta}
\end{equation}
where we adopt $\pi_s=0.14\,\mas$ for the source parallax.

As usual, we must consider the ``ecliptic degeneracy'', which
takes
$(u_0,\alpha,\pi_{\e,\perp},d\alpha/dt)\rightarrow -
(u_0,\alpha,\pi_{\e,\perp},d\alpha/dt)$ \citep{ob09020}, where
$\pi_{\e,\perp}$ is the component of $\bpi_\e$ that is
perpendicular to the direction of Earth's acceleration (projected
on the sky) at $t_0$.  Indeed, because the event is very close
to the ecliptic, $\beta_{\rm ecliptic} = +1.3^\circ$, we expect this
degeneracy to be very severe: it is exact in the limit
$\beta_{\rm ecliptic} \rightarrow 0$.

\begin{figure}
\plotone{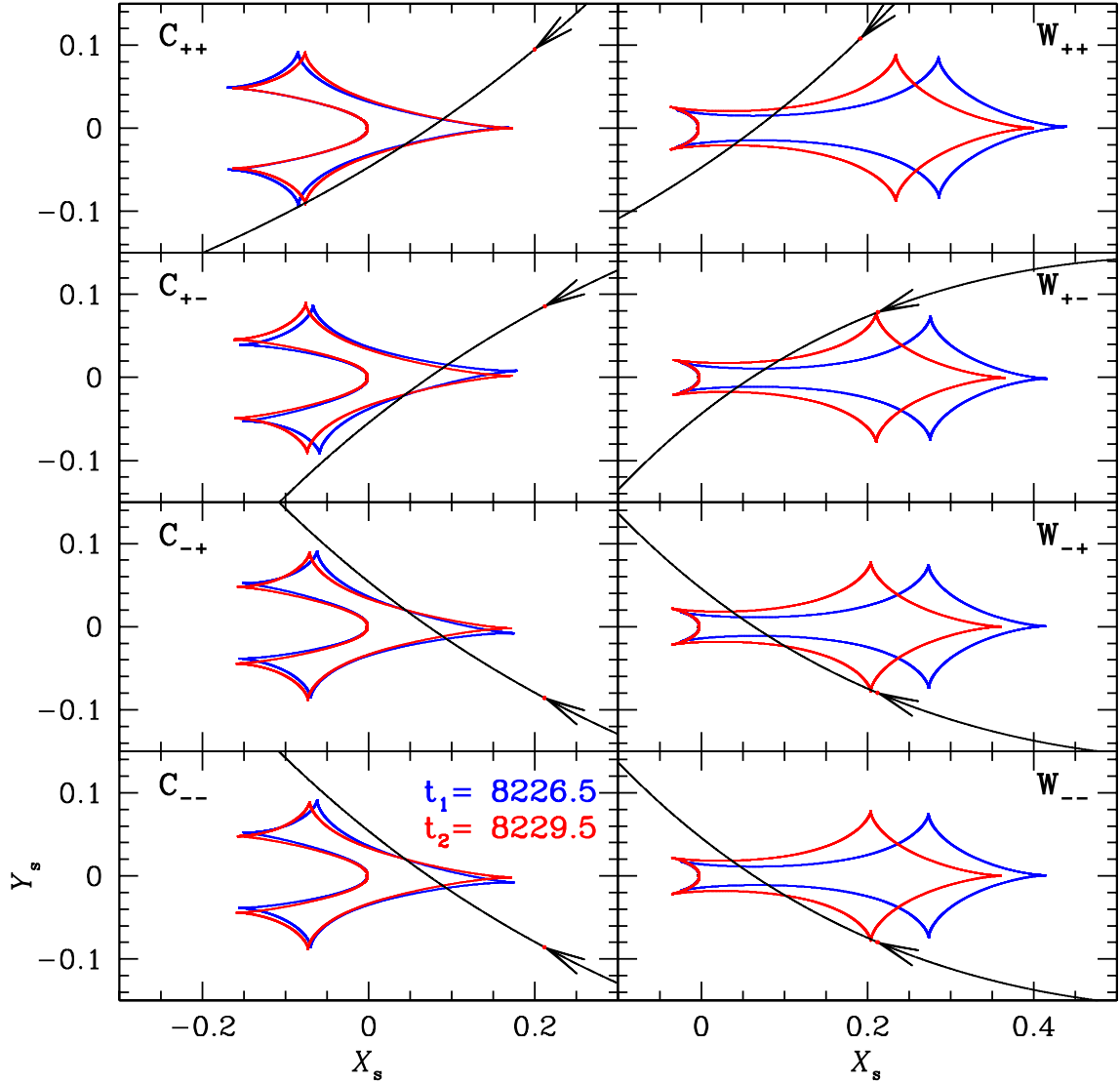}
\caption{Caustic geometries for each of the eight solutions shown
in Tables~\ref{tab:close} and \ref{tab:wide}.  The red and blue closed
curves show the caustics at the two designated epochs, while the
black curves show the source trajectory relative to these structures.
}
\label{fig:geom}
\end{figure}

We find that for each of the four near-degenerate combinations
$[(s<1),(s>1)]\times[(u_0<0),(u_0>0)]$, there are two local
minima, which are characterized by different values of $\pi_{\e,N}$.
Such a discrete degeneracy is predicted by the ``jerk-parallax degeneracy''
\citep{gould04}, according to which
\begin{equation}
\pi_{\e,\perp}^\prime = -(\pi_{\e,\perp} +\pi_{j,\perp})
\label{eqn:jerk}
\end{equation}
where $\pi_{j,\perp}$ is the ``jerk parallax''.  From Equations~(8) and (9)
of \citet{mb03037}, which make a number of simplifying approximations,
\begin{equation}
\pi_{j,\perp} = -{4\over 3}\,{{\rm yr}\over 2\pi t_\e}\,
{\sin\beta_{\rm ecliptic}\over (\cos^2\psi\sin^2\beta_{\rm ecliptic}
+ \sin^2\psi)^{3/2}} \rightarrow -0.05,
\label{eqn:piejperp}
\end{equation}
where $\psi\rightarrow 57^\circ$ is the
ecliptic-latitude offset between the anti-Sun and the event at $t_0$.
Because Earth's projected acceleration is within $0.2^\circ$ of due west
at the peak of the event, $\pi_{\e,\perp} \simeq \pi_{\e,N}$. (Note that
because the $(\pi_{\e,\parallel},\pi_{\e,\perp})$ system is right-handed,
$\pi_{\e,\perp} \sim \pi_{\e,N}$ for the first several months of the
microlensing season and
$\pi_{\e,\perp} \sim -\pi_{\e,N}$ during the last several months.
See, e.g., Figure~3 from \citealt{mb03037}.)
Therefore, from the fact that one $(s<0,u_0<0)$ solution has
$\pi_{\e,N}=1.39$, the jerk-parallax formalism predicts that
the other should have
$\pi_{\e,N}^\prime = -(\pi_{\e,N}+\pi_{j,\perp}) = -1.34$.  The actual value
($-1.12$) is in reasonable agreement, given the error bars.  The
situation is qualitatively similar for the other three pairs of
solutions.  We note
that the jerk-parallax degeneracy has never previously been investigated
in the context of planetary microlensing events.  The caustic geometries
of these eight solutions are shown in Figure~\ref{fig:geom}, and their
parameters are given in Tables~\ref{tab:close} and \ref{tab:wide}.

The first point to note about these eight solutions is that they
have similar $\chi^2$, which means that all must be considered
as viable.  To facilitate the further discussion of these eight solutions,
we label them by, $C_{\pm\pm}$ or $W_{\pm\pm}$.  The letter stands for
``close'' or ``wide'' (for $s<1$ or $s>1$), while the two subscripts
refer to the signs of $u_0$ and $\pi_{\e,N}$, respectively.

Second, we note that six of the eight solutions have very comparable
values of $q$: the close solutions have $q\simeq 3.6\times 10^{-3}$,
while the $W_{+-}$ and $W_{-+}$ solutions have $q\simeq 4.2\times 10^{-3}$.
On the other hand, the solutions $W_{++}$ and $W_{--}$
have significantly higher $q\simeq 5.6\times 10^{-3}$.  Note that
these two solutions have $\Delta\chi^2\sim 4$ relative to the
best solution, and therefore this higher mass ratio will end up
getting less weight in the final parameter estimates.

\begin{figure}
\plotone{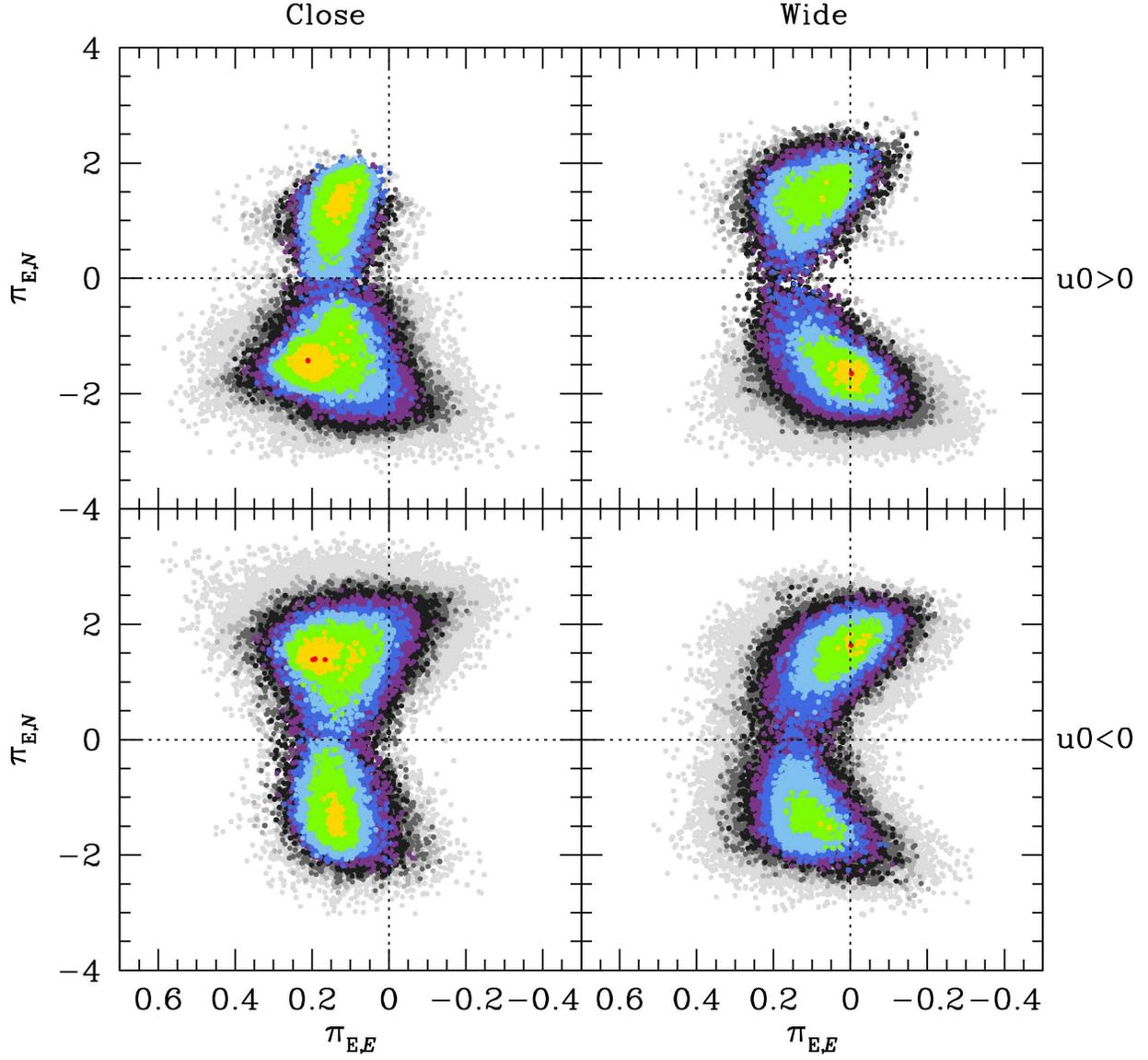}
\caption{Scatter plots from the MCMC of the four solutions
(close,wide)$\times(u_0<0,u_0>0)$ of KMT-2018-BLG-1990. Each shows
two distinct minima, which correspond to the jerk-parallax degeneracy.
Models with $\Delta\chi^2<(1^2,2^2, \ldots 10^2)$ are shown
in (red, gold, green, cyan, blue, magenta, black, dark gray, medium gray,
light gray), respectively.  Note that the abscissa and ordinate axis
scales are incommensurate.
}
\label{fig:bpie}
\end{figure}

Third, all eight solutions have qualitatively comparable values of
$\pi_\e$.  That is, in all cases $|\pi_{\e,N}| \gg |\pi_{\e,E}|$ and
in all cases $|\pi_{\e,N}|$ have comparable values.  These facts
together mean that $\pi_\e\simeq |\pi_{\e,N}|$ are similar for all
four solutions.  Because the errors on $\pi_{\e,N}$ are relatively
large, it is important to assess whether these solutions are well
localized.  We display the $\bpi_\e$ distributions in Figure~\ref{fig:bpie},
which shows that they are indeed well-localized at the two-sigma level,
but less so at the three-sigma level.

Fourth, we note that the remaining microlensing parameters are
also comparable between solutions, with the exception of the
orbital parameters.  However, these are basically just nuisance
parameters, which are very poorly constrained and are included
only to avoid biasing the parallax parameters.

\section{{Physical Parameters}
\label{sec:physical}}

Whenever $\theta_\e$ and $\pi_\e$ are well measured, one can always determine
the relative parallax $\pi_\rel$ and the lens mass $M$ from
Equation~(\ref{eqn:massdist}), and so the lens distance
$D_L = \au/(\theta_\e\pi_\e + \pi_s)$.
In the present case, there are eight different solutions, but the
best fit values for each of these will lead to fairly similar values
$M\sim 0.09\,M_\odot$, $D_L\sim 0.8\,\kpc$.  This distance is unusually
small because the phase space, which grows quadratically with distance,
generally favors more distant lenses.  Smaller values of the microlens
parallax would, by Equation~(\ref{eqn:massdist}), give larger distances.
Therefore, symmetric errors in $\pi_\e$ (which is essentially equal to
$|\pi_{\e,N}|$ in the present case)
should give rise to an asymmetric distribution in
inferred distance around the value derived from the best fit.
When the errors in $\pi_\e$ are small, this effect is likewise small
and can be ignored.  However, in the present case, both the variation
in $\pi_\e$ between solutions and the statistical errors within
solutions are fairly large.  Hence, this effect must be accounted for.
We will address this in Section~\ref{sec:bayes}.  Before doing
so, however, we must first evaluate $\theta_*$ (Section~\ref{sec:cmd})
and investigate the proper-motion distribution of the microlensed
source (Section~\ref{sec:sourcepm}).

\subsection{{CMD}
\label{sec:cmd}}

The first step toward estimating physical parameters is to measure
the source position on the CMD relative to the centroid of the red clump,
which then permits one to estimate $\theta_*$ and thus
$\theta_\e = \theta_*/\rho$.  We first note that while the source
color estimate does not depend on the microlensing model (and
can often be derived by regression of $V$ on $I$ flux,
without any model), the source magnitude does depend on the model.
Therefore, for simplicity, we will explicitly derive results
for the $C_{-+}$ model and then present scaling relations for the
rest.

\begin{figure}
\plotone{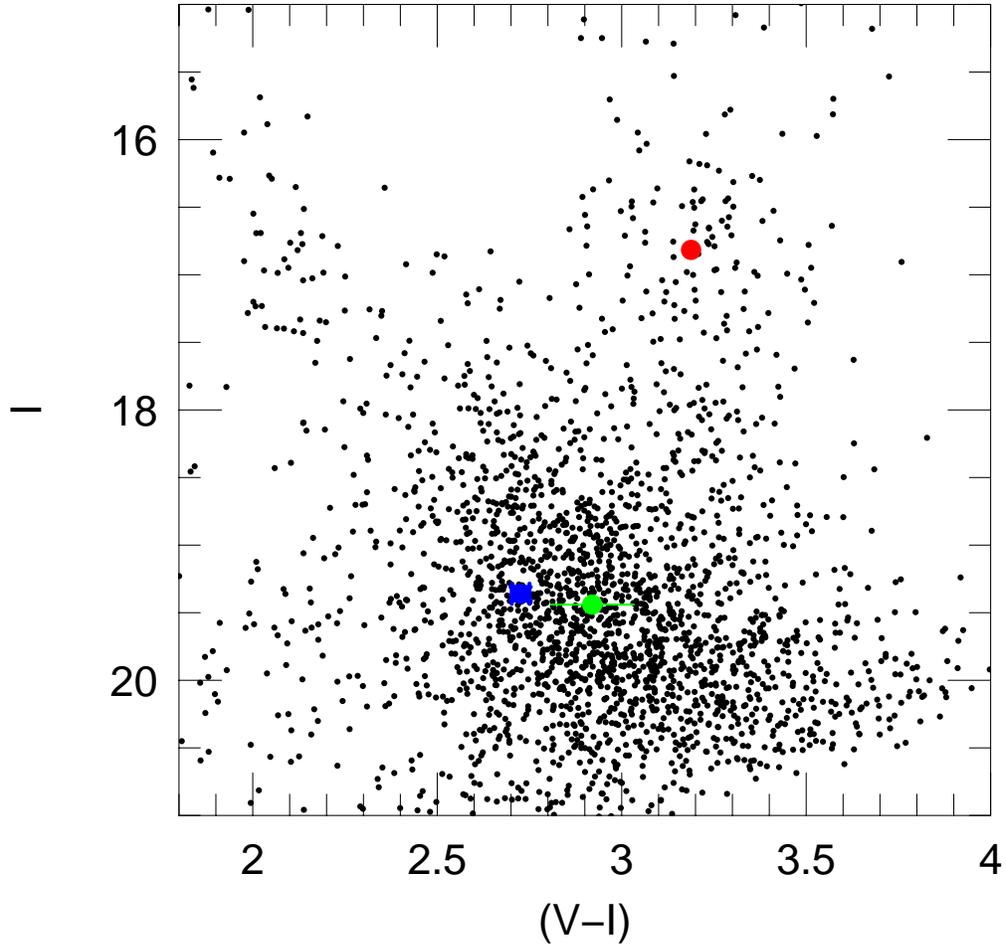}
\caption{Color-Magnitude Diagram (CMD) of stars within a $2^\prime$ square
box centered on KMT-2018-BLG-1990.  The position of the microlensed source,
the blended light and the centroid of the red clump are shown as
blue, green, and red circles, respectively.
The values shown here are taken from the $C_{-+}$ solution, but the
other solutions would be virtually indistinguishable on this plot.
}
\label{fig:cmd}
\end{figure}

The CMD is shown in Figure~\ref{fig:cmd}, with the source and
blend positions marked.  The clump centroid is at
$[(V-I),I]_{\rm cl} = (3.19,16.82)\pm (0.02,0.08)$, whereas the source is at
$[(V-I),I]_{s} = (2.73,19.36)\pm (0.02,0.05)$,
The errors in the clump position are due to centroiding.  The source
color and error are derived from regression without reference to
any model.  The source magnitude error is set to 0.05
as representative of all solutions, which have very similar error bars.
Hence, the offset from the clump is
$\Delta[(V-I),I] = (-0.46,2.54)\pm (0.03,0.10)$.
We adopt
$[(V-I),I]_{\rm cl,0} = (1.06,14.25)$ from \citet{bensby13} and \citet{nataf13},
and so derive
$[(V-I),I]_{s,0} = (0.60,16.79)\pm (0.03,0.10)$.
We have not included any error in the de-reddened position of the clump, but
rather add an overall 5\% error to $\theta_*$ due to all aspects of the method.
We convert from $V/I$ to $V/K$ using the color-color relations of
\citet{bb88}, and then use the color/surface-brightness relation of
\citet{kervella04} to derive
\begin{equation}
\theta_* = 1.23 \pm 0.09 \ \muas;
\qquad
\theta_\e = 0.88\pm 0.11 \ \mas;
\qquad
\mu_\rel = 7.1\pm 0.9\ \masyr,
\label{eqn:thetastar}
\end{equation}
where we have used $\rho=1.39\pm 0.16$ from the $C_{-+}$ solution, and
where we have taken account of the anti-correlation between $f_s$ and $\rho$.
For other solutions, one can simply scale
$\theta_*/\theta_{*,C,-,+}=\sqrt{f_s/0.347}$ and
$\theta_\e/\theta_{\e,C,-,+}=\sqrt{f_s/0.347}/(\rho/1.39\times 10^{-3})$.

\subsection{{Source Proper Motion}
\label{sec:sourcepm}}

The source proper motion $\bmu_s$ is one of many inputs into a Bayesian
estimate of the lens properties.  For relatively bright sources, it
is often possible to measure their proper motions, either from {\it Gaia}
or from ground-based data.  When this is not possible, one usually
adopts an assumed distribution of source proper motions derived from
a Galactic model.  In the present case, the source is too faint and blended
to measure its proper motion from ground data.  In any case, the KMT time
baseline would be too short for an accurate measurement.  Moreover, the
source does not appear in {\it Gaia}.

However, we can still use {\it Gaia} to measure the proper-motion distribution
of ``bulge'' (really, ``bar'') stars, of which the microlensed source is
very likely a member.  We examine a {\it Gaia} CMD and on this basis
select the 82 stars within $1^\prime$ and satisfying $G<18.6$ and $B_p-R_p>2.6$.
We eliminate one outlier and derive (in the Sun frame)
\begin{equation}
\langle\bmu_{\rm bar}(l,b)\rangle = (-5.4,0.1)\pm(0.3,0.2)\,\masyr;
\quad
\sigma(\bmu_{\rm bar}) = (2.9,2.2)\pm(0.2,0.2)\,\masyr
\label{eqn:gaia}
\end{equation}

\subsection{{Bayesian Analysis}
\label{sec:bayes}}

Normally, one does not apply a Bayesian analysis when $\theta_\e$
and $\bpi_\e$ are relatively well measured.  Rather one would
simply combine these measurements using Equation~(\ref{eqn:massdist})
to obtain $M$ and $\pi_\rel$ (and so $D_L$), and propagate the errors
in the measured quantities to the physical quantities.
However, we argued at the beginning of Section~\ref{sec:physical}
that in the present case, a full Bayesian analysis was justified.
Stating the argument differently, if the errors are ``small'',
then the frequentist and
Bayesian approaches will yield nearly identical results, so there
is no point employing the more cumbersome Bayesian formalism.  And whether
the errors can be considered ``small'' depends on the product of their
absolute size and the gradient of the priors.  In the present case,
the errors should be considered ``large'' given the steepness of the
gradient.

We draw random events from a Galactic model, as described in detail
by \citet{ob171522}.  The only exception is that we
draw the source proper motions
from a Gaussian distribution with the parameters that were derived from
Gaia data in Section~\ref{sec:sourcepm}.

Each of the eight solutions is fed exactly the same ensemble of $10^8$
simulated events.  For each solution, each simulated event is given
a weight equal to the likelihood of its four inferred parameters
$(t_\e,\theta_\e,\pi_{\e,N},\pi_{\e,E})$ given the error distributions of
these quantities derived from the MCMC for that
solution.  Finally, we weight by $\exp(-\Delta\chi^2_k/2)$,
where $\Delta\chi^2_k$ is the difference in $\chi^2$ between the $k$th
solution and the best fit.

Hence, the $i$th simulated event in the $k$th solution
is given a weight of
\begin{equation}
w_{i,k} = {\cal L}_{i,k}(t_\e) {\cal L}_{i,k}(\theta_\e) {\cal L}_{i,k}(\bpi_\e)
\exp(-\Delta\chi^2_k/2)
\label{eqn:weight}
\end{equation}
where
\begin{equation}
{\cal L}_{i,k}(X) ={\exp[-(X_i-X_k)^2/2\sigma_{X_k}^2]
\over 2\pi\sigma_{X_k}},
\qquad
X=t_\e\ {\rm or}\ X=\theta_\e,
\label{eqn:likes}
\end{equation}
\begin{equation}
{\cal L}_{i,k}(\bpi_\e)=
{\exp[-\sum_{m,n=1}^2 b^k_{m,n} (\pi_{\e,m,i}-\pi_{\e,m,k})(\pi_{\e,n,i}-\pi_{\e,n,k})/2]
\over (2\pi)^2/\sqrt{{\rm det}\, b^k}},
\label{eqn:likes2}
\end{equation}
$b^k_{m,n}$ is the inverse covariance matrix of the $\bpi_\e$ parallax vector
in the $k$-th solution, and $(m,n)$ are dummy variables ranging over
$(N,E)$.

\begin{figure}
\plotone{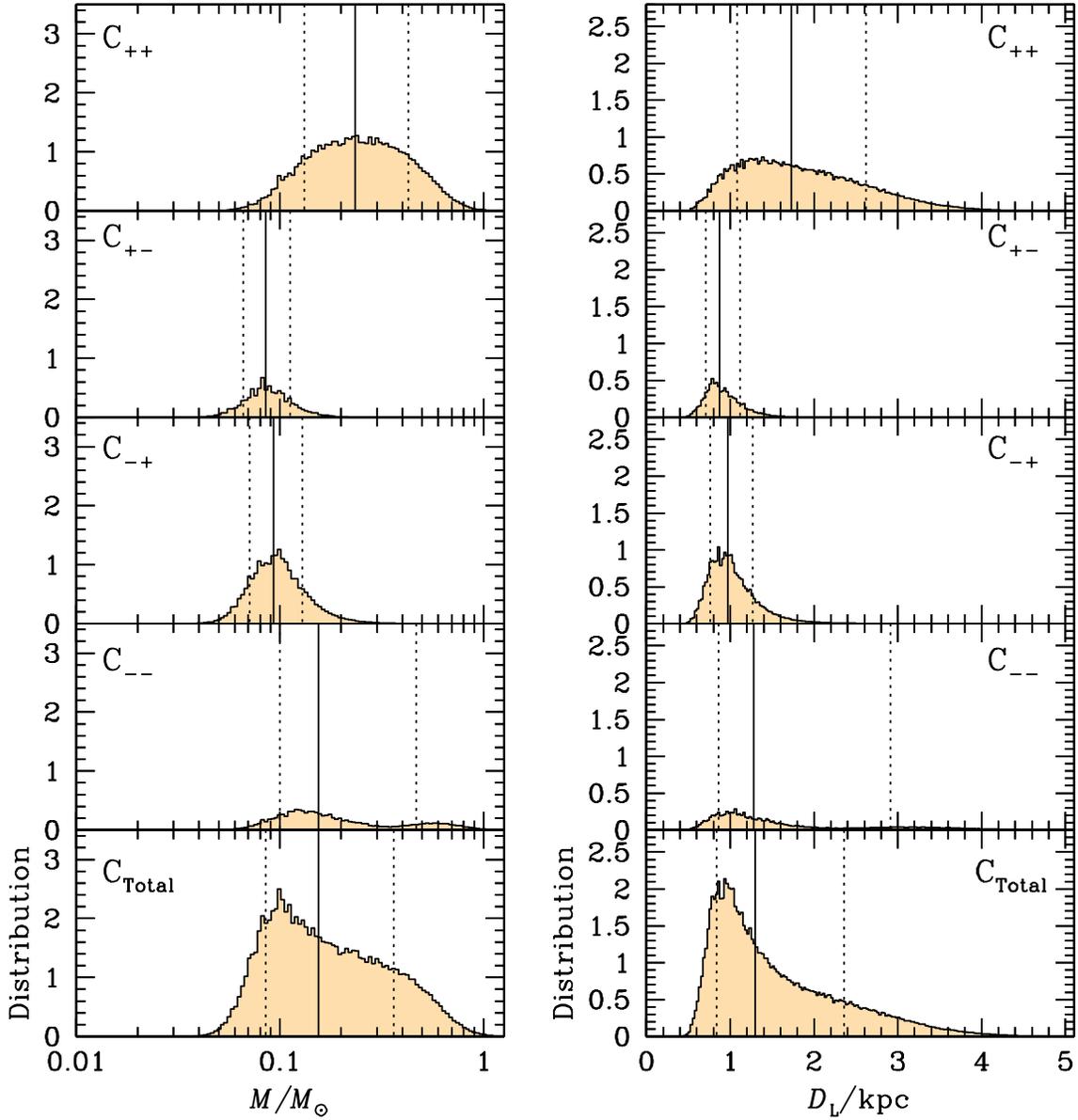}
\caption{Histogram of the Bayesian posteriors of the lens mass and distance
for each of the four of close solutions (top four rows) and the sum of
all four solutions (bottom row).  The total area in each histogram
is proportional to the product of the weight derived from Bayesian
analysis and $\exp(-\Delta\chi^2/2)$ of the solution.  The $C_{++}$
solution ($u_0>0$, $\pi_{\e,N}>0$)
dominates primarily because it is compatible with more distant
lenses, which are more numerous in the Galactic model.
}
\label{fig:hist_close}
\end{figure}

\begin{figure}
\plotone{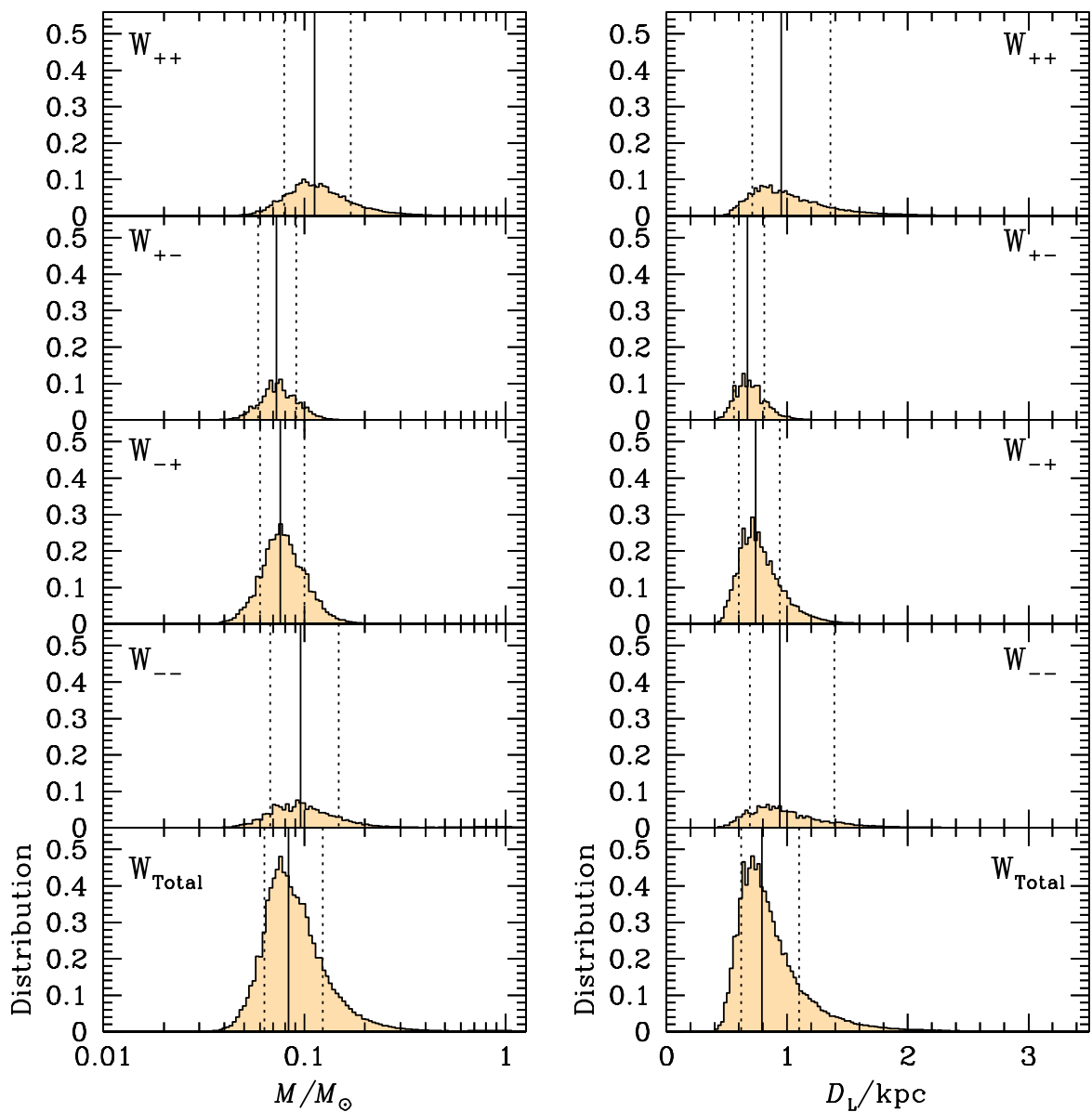}
\caption{Histogram of wide solutions.  The plot is similar to
Figure~\ref{fig:hist_close}.  However, the overall scale is lower
by a factor of roughly five because these solutions are less
favored by both $\chi^2$ and the Galactic model.
}
\label{fig:hist_wide}
\end{figure}

The upper four sets of panels in Figures~\ref{fig:hist_close} and
\ref{fig:hist_wide} show histograms for the lens mass and distance
for each of the four close solutions and each of the four wide solutions,
respectively.  In each case, the relative area under the curve is the
result of the weightings described by
Equations~(\ref{eqn:weight})--(\ref{eqn:likes2}).
The relative weights can be evaluated from the last two columns of
Table~\ref{tab:phys}.
In the bottom two panels of each figure, we show the combined histograms.

\begin{figure}
\plotone{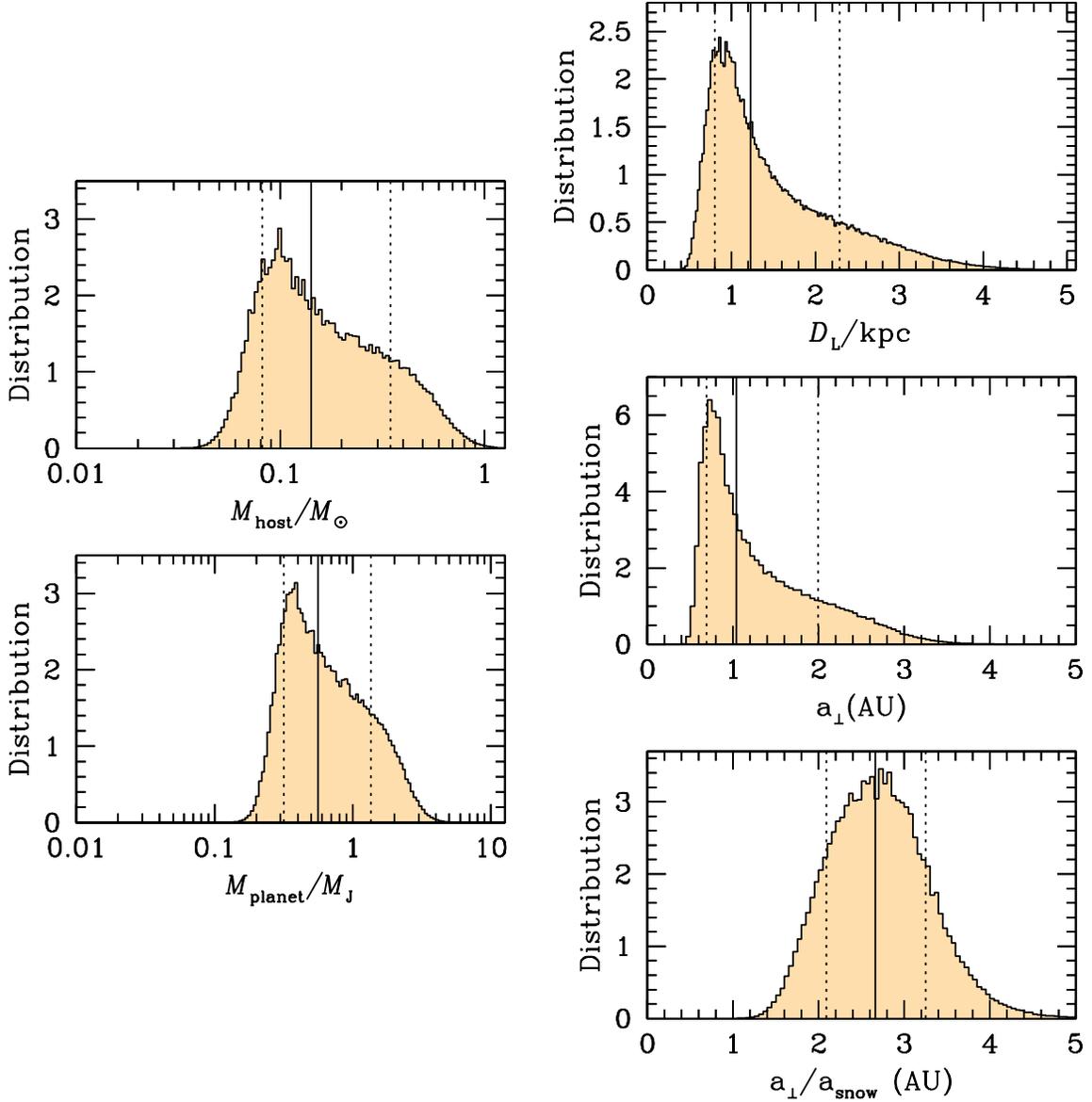}
\caption{Posterior distributions of five physical quantities:
host mass, planet mass, system distance, plant-host projected
separation, and projected separation normalized to the snow line.
Despite the parallax measurement, the allowed mass range in quite
broad.  Unless the lens proves to be substellar (for which there
is only a small probability), its mass and distance will be
decisively resolved at first light of next-generation (30m) telescopes.
}
\label{fig:hist_total}
\end{figure}

The first point to note is that, overall, the close solutions are
significantly favored.  This is mainly due to the fact that the
$C_{++}$ solution has
the highest combined weight, which in turn reflects that it is
compatible with disk lenses at relatively large distances (1--3 kpc),
where the observation cone contains substantially more stars.
The relatively broad range of lens distances of the dominant $C_{++}$ solution
(together with the well constrained $\theta_\e=\sqrt{\kappa M \pi_\rel}$),
likewise implies a relative broad distribution of lens masses.
We conclude that the lens is very likely to be a low-mass star,
within a factor of two of $M=0.1.5\,M_\odot$, at a distance of 0.8--2.3 kpc.
See Figure~\ref{fig:hist_total}.
Because both the ``wide'' and ``close'' solutions have
normalized separation $s$ quite close
to unity, they do not substantially differ in their implications
for the host-planet projected separation $a_\perp\equiv s\theta_\e D_L$:
the uncertainty in this quantity is dominated by the uncertain source
distance.  This uncertainty mainly factors out if we normalize
this projected separation to the snow line, for which we adopt
$a_{\rm snow} = 2.7\au(M/M_\odot)$,
\begin{equation}
{a_\perp\over a_{\rm snow}} = {\kappa M_\odot s/2.7\over \theta_\e + \pi_s/\pi_\e}
\simeq {(3\,\mas)s \over \theta_\e + \pi_s/\pi_\e}
\label{eqn:snowlinerat}
\end{equation}
Thus, for very nearby lenses, i.e., $\pi_\e\gg \pi_s$,
$a_\perp/a_{\rm snow}\rightarrow 3.4$, whereas for relatively distant
lenses (within the framework of the Bayesian posteriors), $D_L\sim 3\,\kpc$,
$a_\perp/a_{\rm snow}\rightarrow 2.0$.  That is, the lens lies projected
well outside the snow line over almost all of the posterior probability
distribution.


\section{{Discussion}
\label{sec:discuss}}

Despite a reasonably strong microlens parallax signal, the mass
and distance of KMT-2018-BLG-1990 remain uncertain at the factor two
level.  Nevertheless, a Bayesian analysis combined with the parallax
measurement shows that the lens is very likely to be a low mass M dwarf
orbited by a gas giant planet in the Saturn-to-Jupiter mass range.

The relatively high lens-source relative proper motion
$\mu_\rel \simeq 7\,\masyr$ implies that by 2028 (i.e., the roughly
expected adaptive-optics (AO) first light on next-generation 30m telescopes),
the lens and source will be separated by about 70 mas, which is easily
enough to separately resolve them.  At that time the remaining uncertainty
in the lens mass and distance can be resolved.  According to the
Bayesian analysis presented in Section~\ref{sec:bayes}, there is
a small chance that the lens is below the hydrogen-burning limit,
in which case it would most likely not be detected in such AO follow-up
observation.  Even in this case, however, it would be known to be
a substellar object.

This is the second KMT-only Jovian planet that has been detected in the
$44\,\rm deg^2$ of low-cadence $\Gamma\sim 0.4\,{\rm hr}^{-1}$ KMTNet
observations.  This confirms the naive expectation,
which we outlined in Section~\ref{sec:intro}, that KMTNet's
3-site survey at this cadence should be sensitive to such planets.
Improving statistics on this population is crucial to our understanding
of planet formation and early orbital evolution of planetary systems.
Guided in part by the observed planet-mass distribution in the
Solar System and in part by theoretical consideration, it has long
been predicted that the outer parts of extra-solar planetary systems
should show a ``gap'' between Neptune-mass and Jovian-mass planets.
Yet, the analysis of microlensing surveys (the only available means
at present to probe cold planets that may lie in this gap) do not
confirm these expectations.  By obtaining a substantial sample of
planets in this range, one can lay the basis for measuring their
masses in subsequent AO followup observations, thus permitting
a much stronger test of the ``gap'' hypothesis.

\begin{equation}
\label{eqn:}
\end{equation}

\acknowledgments
Work by AG was supported by AST-1516842 from the US NSF.
IGS and AG were supported by JPL grant 1500811.
AG received support from the European  Research  Council  under  the  European  Union’s Seventh Framework Programme (FP 7) ERC Grant Agreement n. [321035].
Work by CH was supported by the grant (2017R1A4A1015178) of National
Research Foundation of Korea.
This research has made use of the KMTNet system operated by the Korea
Astronomy and Space Science Institute (KASI) and the data were obtained at
three host sites of CTIO in Chile, SAAO in South Africa, and SSO in
Australia.
\begin{deluxetable}{lcc}
\tablecolumns{3} \tablewidth{0pc} \tablecaption{\textsc{Standard
models}} \tablehead{ \colhead{} & \colhead{Close} & \colhead{Wide}}
\startdata
  $\chi^2/\rm{dof}$               &786.061/749           &788.832/749          \\
  $t_0$ $(\rm{HJD}^{\prime})$     &8230.495 $\pm$ 0.019  &8230.370 $\pm$ 0.021 \\
  $u_0$                           &0.044 $\pm$ 0.001     &0.039 $\pm$ 0.001    \\
  $t_{\rm E}$ $(\rm{days})$       &43.929 $\pm$ 0.754    &45.299 $\pm$ 0.984   \\
  $s$                             &0.968 $\pm$ 0.001     &1.146 $\pm$ 0.004    \\
  $q$ $(10^{-3})$                 &3.723 $\pm$ 0.167     &6.229 $\pm$ 0.315    \\
  $\alpha$ $(\rm{rad})$           &2.559 $\pm$ 0.007     &2.482 $\pm$ 0.007    \\
  $\rho$ $(10^{-3})$              &1.280 $\pm$ 0.144     &1.260 $\pm$ 0.155    \\
  $f_S$                           &0.356 $\pm$ 0.008     &0.344 $\pm$ 0.010    \\
  $f_B$                           &0.147 $\pm$ 0.007     &0.158 $\pm$ 0.008    \\
  $t_*$ $(\rm{days})$             &0.056 $\pm$ 0.006     &0.057 $\pm$ 0.007    \\
\enddata
\tablecomments{Zerpoint for fluxes is 18, e.g., $I_S = 18 -
2.5\log(f_S)$.} \label{tab:ulens}
\end{deluxetable}

\begin{deluxetable}{lccccc}
\tablecolumns{6} \tablewidth{0pc}
\tablecaption{\textsc{Parallax+orbital motion models for the
close-separation}} \tablehead{ \colhead{} &
\multicolumn{2}{c}{$u_0>0$} & \colhead{} &
\multicolumn{2}{c}{$u_0<0$}\\
\cline{2-3} \cline{5-6} \colhead{Parameters} &
\colhead{$\pi_{\rm{E},\it{N}}>0$} &
\colhead{$\pi_{\rm{E},\it{N}}<0$} & \colhead{} &
\colhead{$\pi_{\rm{E},\it{N}}>0$} &
\colhead{$\pi_{\rm{E},\it{N}}<0$}} \startdata
  $\chi^2/\rm{dof}$               &743.899/745           &742.074/745          & &742.013/745           &743.720/745           \\
  $t_0$ $(\rm{HJD}^{\prime})$     &8230.449 $\pm$ 0.024  &8230.466 $\pm$ 0.024 & &8230.465 $\pm$ 0.024  &8230.448 $\pm$ 0.023  \\
  $u_0$                           &0.040 $\pm$ 0.002     &0.043 $\pm$ 0.002    & &-0.043 $\pm$ 0.002    &-0.040 $\pm$ 0.002    \\
  $t_{\rm E}$ $(\rm{days})$       &46.484 $\pm$ 1.945    &45.372 $\pm$ 1.760   & &45.908 $\pm$ 2.228    &46.223 $\pm$ 2.260    \\
  $s$                             &0.963 $\pm$ 0.003     &0.960 $\pm$ 0.007    & &0.963 $\pm$ 0.007     &0.963 $\pm$ 0.005     \\
  $q$ $(10^{-3})$                 &3.748 $\pm$ 0.253     &3.648 $\pm$ 0.370    & &3.529 $\pm$ 0.359     &3.774 $\pm$ 0.303     \\
  $\alpha$ $(\rm{rad})$           &2.584 $\pm$ 0.018     &2.488 $\pm$ 0.017    & &3.794 $\pm$ 0.021     &3.699 $\pm$ 0.020     \\
  $\rho$ $(10^{-3})$              &1.187 $\pm$ 0.162     &1.423 $\pm$ 0.150    & &1.388 $\pm$ 0.157     &1.228 $\pm$ 0.159     \\
  $\pi_{\rm{E},\it{N}}$           &1.147 $\pm$ 0.349     &-1.428 $\pm$ 0.273   & &1.386 $\pm$ 0.262     &-1.122 $\pm$ 0.355    \\
  $\pi_{\rm{E},\it{E}}$           &0.140 $\pm$ 0.031     &0.211 $\pm$ 0.046    & &0.198 $\pm$ 0.049     &0.146 $\pm$ 0.034     \\
  $ds/dt$ $(\rm{yr}^{-1})$        &0.503 $\pm$ 0.232     &-0.695 $\pm$ 0.574   & &-0.359 $\pm$ 0.539    &0.509 $\pm$ 0.548     \\
  $d\alpha/dt$ $(\rm{yr}^{-1})$   &-0.302 $\pm$ 1.197    &-4.136 $\pm$ 1.764   & &4.139 $\pm$ 2.294     &0.163 $\pm$ 1.244     \\
  $f_S$                           &0.325 $\pm$ 0.016     &0.353 $\pm$ 0.015    & &0.347 $\pm$ 0.018     &0.328 $\pm$ 0.017     \\
  $f_B$                           &0.176 $\pm$ 0.016     &0.151 $\pm$ 0.014    & &0.154 $\pm$ 0.017     &0.175 $\pm$ 0.017     \\
  $t_*$ $(\rm{days})$             &0.055 $\pm$ 0.007     &0.065 $\pm$ 0.007    & &0.064 $\pm$ 0.007     &0.057 $\pm$ 0.007     \\
  $\beta$                         &0.018 $\pm$ 0.111     &0.653 $\pm$ 0.217    & &0.666 $\pm$ 0.201     &0.016 $\pm$ 0.154     \\
\enddata
\tablecomments{The parameter $\beta$ is restricted to $\beta<0.8$.}
\label{tab:close}
\end{deluxetable}

\begin{deluxetable}{lccccc}
\tablecolumns{6} \tablewidth{0pc}
\tablecaption{\textsc{Parallax+orbital motion models for the
wide-separation}} \tablehead{ \colhead{} &
\multicolumn{2}{c}{$u_0>0$} & \colhead{} &
\multicolumn{2}{c}{$u_0<0$}\\
\cline{2-3} \cline{5-6} \colhead{Parameters} &
\colhead{$\pi_{\rm{E},\it{N}}>0$} &
\colhead{$\pi_{\rm{E},\it{N}}<0$} & \colhead{} &
\colhead{$\pi_{\rm{E},\it{N}}>0$} &
\colhead{$\pi_{\rm{E},\it{N}}<0$}} \startdata
  $\chi^2/\rm{dof}$               &746.620/745           &743.078/745          & &742.497/745           &745.785/745           \\
  $t_0$ $(\rm{HJD}^{\prime})$     &8230.343 $\pm$ 0.030  &8230.405 $\pm$ 0.025 & &8230.400 $\pm$ 0.027  &8230.327 $\pm$ 0.024  \\
  $u_0$                           &0.039 $\pm$ 0.002     &0.035 $\pm$ 0.002    & &-0.037 $\pm$ 0.002    &-0.039 $\pm$ 0.002    \\
  $t_{\rm E}$ $(\rm{days})$       &45.646 $\pm$ 1.833    &53.380 $\pm$ 2.721   & &51.389 $\pm$ 3.152    &46.228 $\pm$ 1.870    \\
  $s$                             &1.115 $\pm$ 0.012     &1.099 $\pm$ 0.009    & &1.095 $\pm$ 0.010     &1.118 $\pm$ 0.009     \\
  $q$ $(10^{-3})$                 &5.591 $\pm$ 0.432     &4.171 $\pm$ 0.418    & &4.185 $\pm$ 0.485     &5.695 $\pm$ 0.335     \\
  $\alpha$ $(\rm{rad})$           &2.530 $\pm$ 0.015     &2.504 $\pm$ 0.011    & &3.781 $\pm$ 0.016     &3.750 $\pm$ 0.015     \\
  $\rho$ $(10^{-3})$              &1.223 $\pm$ 0.158     &1.175 $\pm$ 0.140    & &1.246 $\pm$ 0.155     &1.336 $\pm$ 0.160     \\
  $\pi_{\rm{E},\it{N}}$           &1.400 $\pm$ 0.287     &-1.649 $\pm$ 0.208   & &1.637 $\pm$ 0.252     &-1.521 $\pm$ 0.370    \\
  $\pi_{\rm{E},\it{E}}$           &0.075 $\pm$ 0.047     &-0.003 $\pm$ 0.038   & &-0.002 $\pm$ 0.043    &0.056 $\pm$ 0.034     \\
  $ds/dt$ $(\rm{yr}^{-1})$        &-3.580 $\pm$ 1.249    &-4.464 $\pm$ 0.727   & &-4.762 $\pm$ 0.098    &-3.429 $\pm$ 0.855    \\
  $d\alpha/dt$ $(\rm{yr}^{-1})$   &-0.367 $\pm$ 1.397    &0.288 $\pm$ 0.665    & &-0.156 $\pm$ 1.269    &-0.033 $\pm$ 0.348    \\
  $f_S$                           &0.334 $\pm$ 0.016     &0.294 $\pm$ 0.017    & &0.307 $\pm$ 0.020     &0.329 $\pm$ 0.017     \\
  $f_B$                           &0.169 $\pm$ 0.016     &0.209 $\pm$ 0.017    & &0.195 $\pm$ 0.020     &0.175 $\pm$ 0.017     \\
  $t_*$ $(\rm{days})$             &0.056 $\pm$ 0.007     &0.063 $\pm$ 0.007    & &0.064 $\pm$ 0.007     &0.062 $\pm$ 0.007     \\
  $\beta$                         &0.567 $\pm$ 0.187     &0.625 $\pm$ 0.144    & &0.750 $\pm$ 0.187     &0.477 $\pm$ 0.133     \\
\enddata
\tablecomments{The parameter $\beta$ is restricted to $\beta<0.8$.}
\label{tab:wide}
\end{deluxetable}

\begin{deluxetable}{lccccccc}
\tablecolumns{8} \tablewidth{0pc}\tablecaption{\textsc{Physical
properties}} \tablehead{\colhead{} & \multicolumn{4}{c}{Physical
Properties} & \colhead{} &
\multicolumn{2}{c}{Relative Weights}\\
\cline{2-5} \cline{7-8} \colhead{Models}&\colhead{$M_{\rm host}$
$[M_\sun]$} & \colhead{$M_{\rm planet}$ $[M_J]$} & \colhead{$D_{\rm
L}$ [kpc]} & \colhead{$a_\bot$ [au]} & \colhead{} &
\colhead{Gal.Mod.} & \colhead{$\chi^2$}} \startdata
  $C_{++}$  &$0.235_{-0.104}^{+0.187}$  &$0.923_{-0.409}^{+0.734}$ &$1.735_{-0.651}^{+0.889}$  &$1.521_{-0.555}^{+0.753}$ &&1.000 &0.390\\
  $C_{+-}$  &$0.084_{-0.019}^{+0.026}$  &$0.323_{-0.071}^{+0.100}$ &$0.873_{-0.165}^{+0.249}$  &$0.689_{-0.103}^{+0.171}$ &&0.067 &0.970\\
  $C_{-+}$  &$0.094_{-0.024}^{+0.036}$  &$0.348_{-0.087}^{+0.132}$ &$0.967_{-0.212}^{+0.308}$  &$0.763_{-0.145}^{+0.219}$ &&0.171 &1.000\\
  $C_{--}$  &$0.153_{-0.054}^{+0.311}$  &$0.605_{-0.215}^{+1.229}$ &$1.277_{-0.422}^{+1.629}$  &$1.083_{-0.315}^{+1.353}$ &&0.189 &0.425\\
  $C_{\rm Total}$  &$0.153_{-0.068}^{+0.206}$  &$0.594_{-0.272}^{+0.814}$ &$1.296_{-0.452}^{+1.061}$  &$1.095_{-0.391}^{+0.953}$ && &\\
  \cline{1-8}
  $W_{++}$  &$0.111_{-0.031}^{+0.057}$  &$0.649_{-0.181}^{+0.330}$ &$0.950_{-0.238}^{+0.411}$  &$0.947_{-0.203}^{+0.377}$ &&0.159 &0.100\\
  $W_{+-}$  &$0.073_{-0.014}^{+0.018}$  &$0.318_{-0.058}^{+0.077}$ &$0.674_{-0.115}^{+0.139}$  &$0.657_{-0.075}^{+0.097}$ &&0.017 &0.587\\
  $W_{-+}$  &$0.076_{-0.016}^{+0.023}$  &$0.333_{-0.068}^{+0.101}$ &$0.742_{-0.138}^{+0.201}$  &$0.699_{-0.107}^{+0.145}$ &&0.039 &0.785\\
  $W_{--}$  &$0.095_{-0.027}^{+0.053}$  &$0.565_{-0.161}^{+0.316}$ &$0.937_{-0.243}^{+0.457}$  &$0.883_{-0.207}^{+0.379}$ &&0.078 &0.152\\
  $W_{\rm Total}$  &$0.083_{-0.020}^{+0.039}$  &$0.398_{-0.112}^{+0.291}$ &$0.789_{-0.169}^{+0.316}$  &$0.751_{-0.137}^{+0.283}$ && &\\
  \cline{1-8}
  ${\rm Total}$  &$0.141_{-0.061}^{+0.202}$  &$0.565_{-0.249}^{+0.789}$ &$1.229_{-0.427}^{+1.063}$  &$1.039_{-0.351}^{+0.951}$ && &\\
\enddata
\label{tab:phys}
\end{deluxetable}
\clearpage

\begin{thebibliography}{99}


\bibitem[Alard \& Lupton(1998)]{alard98} Alard, C. \& Lupton, R.H.,1998, \apj, 503, 325



\bibitem[Albrow et al.(2009)]{albrow09}Albrow, M.\ D., Horne, K., Bramich, D.\ M., et al.\ 2009, \mnras, 397, 2099

\bibitem[Albrow(2017)]{pydia}Albrow, M. D., MichaelDAlbrow/pyDIA: Initial Release on Github, doi:10.5281/zenodo.268049

















\bibitem[Bensby et al.(2013)]{bensby13} Bensby, T. Yee, J.C., Feltzing, S.\ et al.\ 2013, \aap, 549A, 147

\bibitem[Bessell \& Brett(1988)]{bb88} Bessell, M.S., \& Brett, J.M.\ 1988, \pasp, 100, 1134





























\bibitem[Gould(1992)]{gould92} Gould, A. 1992, \apj, 392, 442






\bibitem[Gould(2000)]{gould00} Gould, A. 2000, \apj, 542, 785

\bibitem[Gould(2004)]{gould04} Gould, A. 2004, \apjl, 606, 319







\bibitem[Gould \& Loeb(1992)]{gouldloeb} Gould, A. \& Loeb, A. 1992, \apj, 396, 104






\bibitem[Gould et al.(2013)]{prop2013} Gould, A., Carey, S., \& Yee, J. 2013, 2013spitz.prop.10036

\bibitem[Gould et al.(2014)]{prop2014} Gould, A., Carey, S., \& Yee, J. 2014, 2014spitz.prop.11006

\bibitem[Gould et al.(2015a)]{prop2015a} Gould, A., Yee, J., \& Carey, S., 2015a, 2015spitz.prop.12013

\bibitem[Gould et al.(2015b)]{prop2015b} Gould, A., Yee, J., \& Carey, S. 2015b, 2016spitz.prop.12015

\bibitem[Gould et al.(2016)]{prop2016} Gould, A., Carey, S., \& Yee, J. 2016, 2016spitz.prop.13005

\bibitem[Gould et al.(2018)]{prop2018} Gould, A., Yee, J., Carey, S., \& Shvartzvald, Y.\ 2018, 2018spitz.prop.14012




\bibitem[Griest \& Safizadeh(1998)]{griest98} Griest, K.\ \& Safizadeh, N.\ 1998, \apj, 500, 37






\bibitem[Henderson et al.(2014)]{henderson14} Henderson, C.B., Gaudi, B.S., Han, c., et al. 2014, \apj, 794, 52










\bibitem[Jung et al.(2018)]{ob171522}Jung, Y.\ K., Udalski, A., Gould, A.,  2018 \aj, 155, 219




\bibitem[Kervella et al.(2004)]{kervella04} Kervella, P., Th{\'e}venin, F., Di Folco, E., \& S{\'e}gransan, D.\ 2004, \aap, 426, 297

\bibitem[Kim et al.(2016)]{kmtnet} Kim, S.-L., Lee, C.-U., Park, B.-G., et al.  2016, JKAS, 49, 37

\bibitem[Kim et al.(2018a)]{eventfinder} Kim, D.-J., Kim,  H.-W., Hwang, K.-H., et al., 2018a, \aj, 155, 76


\bibitem[Kim et al.(2018c)]{2016eventfinder} Kim,  H.-W., Hwang, K.-H., Kim, D.-J., et al., 2018c, AAS submitted, arXiv:1804.03352

\bibitem[Kim et al.(2018d)]{alertfinder} Kim,  H.-W., Hwang, K.-H., Shvartzvald, Y. et al., 2018d, AAS submitted, arXiv:1806.07545



\bibitem[Mao \& Paczy\'nski(1991)]{mao91} Mao, S.\ \& Paczy\'nski, B.\ 1991, \apj, 374, 37



\bibitem[Nataf et al.(2013)]{nataf13} Nataf, D.M., Gould, A., Fouqu\'e, P. et al. 2013, \apj, 769, 88


\bibitem[Paczy\'nski(1986)]{pac86} Paczy\'nski, B.\ 1986, \apj, 304, 1

\bibitem[Park et al.(2004)]{mb03037}Park, B.-G., DePoy, D.L.., Gaudi, B.S.,  et al.\ 2004, \apj, 609, 166


































\bibitem[Skowron et al.(2011)]{ob09020}Skowron, J., Udalski, A., Gould, A et al.\ 2011, \apj, 738, 87
















\bibitem[Udalski et al.(2005)]{ob05071} Udalski, A., Jaroszy\'nski, M., Paczy\'nski, B, et al. 2005, \apj, 628, L109.







\bibitem[Wo\'zniak(2000)]{wozniak2000} Wo\'zniak, P.~R. 2000, Acta Astron., 50, 421








\bibitem[Zang et al.(2018)]{kb161397} Zang, W., Hwang, K.-H., Kim, H.-W., et al. 2018, \aj, 156, 236








\end{thebibliography}
\end{document}